\documentclass[superscriptaddress,reprint,amsmath,amssymb,aps,pra,floatfix,twocolumn]{revtex4-2}

\usepackage[T1]{fontenc}
\usepackage{graphicx}
\usepackage{dcolumn}
\usepackage{bm}
\usepackage{xcolor}
\usepackage{etoolbox}
\usepackage[colorlinks = true,linkcolor = blue,urlcolor  = blue,citecolor = blue, anchorcolor = blue]{hyperref}
\usepackage{orcidlink}

\begin{document}
%\preprint{AIP/123-QED}

%\documentclass[a4paper, amsfonts, amssymb, amsmath, reprint, showkeys, nofootinbib, https://www.overleaf.com/project/653b8bcef7c27317048f2ac7twoside]{revtex4-1}
%\usepackage[english]{babel}
%\usepackage[utf8]{inputenc}
%\usepackage[colorinlistoftodos, color=green!40, prependcaption]{todonotes}
%\input{preamble}
%\usepackage[pdftex, pdftitle={Article}, pdfauthor={Author}]{hyperref} % For hyperlinks in the PDF
%\setlength{\marginparwidth}{2.5cm}
%\bibliographystyle{apsrev4-1}
%\begin{document}
\title{Negative Interaction Quench Dynamics of Density-Ordered Dipolar Bosons in a One-Dimensional Optical Lattice}
\author{Rhombik Roy}
\affiliation{Department of Physics, University of Haifa, Haifa 3498838, Israel.}
\affiliation{Haifa Research Center for Theoretical Physics and Astrophysics,
University of Haifa, Haifa 3498838, Israel.}
\author{N.~D.~Chavda \orcidlink{0000-0001-5958-1143}}
\affiliation{Department of Applied Physics, Faculty of Technology and Engineering,\\The Maharaja Sayajirao University of Baroda, Vadodara-390001, India. }
\author{Barnali Chakrabarti\orcidlink{0000-0002-6320-9894}} \email{barnali@if.usp.br}
\affiliation{Departamento de Física,
Universidade Federal de Pernambuco, 50670-901 Recife, Pernambuco, Brazil}
\affiliation{Instituto de Física, Universidade de São Paulo, CEP 05508-090, SP, Brazil.}
\author{Arnaldo Gammal\orcidlink{0000-0003-4720-3203}}
\affiliation{Instituto de Física, Universidade de São Paulo, CEP 05508-090, SP, Brazil.}
%\email{bchakrab@ictp.it}
%\affiliation{$^1$Department of Physics, Presidency University, 86/1   College Street, Kolkata 700073, India.}
%\affiliation{$^2$Instituto de Física, Universidade de São Paulo, CEP 05508-090, SP, Brazil.}
%\affiliation{$^3$Dipartimento di Fisica ``Galileo Galilei", Universita di Padova, INFN Sezione di Padova, and CNR-INO Unita di Sesto Fiorentino, Italy}
\date{\today} % Leave empty to omit a date

\begin{abstract}

We explore the nonequilibrium dynamics of a density-ordered dipolar Bose gas in a finite one-dimensional optical lattice following a negative interaction quench, using the numerically exact multiconfigurational time-dependent Hartree method for bosons. The interaction sign reversal, effectively driving a crossover from long-range to short-range interactions, generates rich intra- and interwell tunneling dynamics spanning superfluid, Mott-insulating, and fragmented regimes. A striking finding is the robustness of the underlying crystal-state correlations against the quench, despite the strong dynamical response. We identify emergent excitation modes, including local breathing and dipole-like oscillations, via real- and momentum-space observables, and quantify tunneling through site-resolved position variance. One- and two-body Glauber correlation functions further uncover a direct connection between tunneling and correlation dynamics. Moreover, we show that combining interaction quenches with lattice-depth ramping enables controllable dynamical engineering, establishing dipolar lattice systems as a promising platform for nonequilibrium quantum simulation.
\end{abstract}

%\keywords{disorder, Mott localization, correlation}

\maketitle

%\section{Introduction} \label{sec:intro}
\section{Introduction}
Ultracold atomic gases in optical lattices provide a highly controllable platform to study nonequilibrium dynamics in strongly correlated quantum systems. Precise control of laser fields and tunable interactions allows manipulation of both the lattice potential and interparticle interactions, making optical lattices ideal for exploring tunneling dynamics, transport phenomena, and the emergence of correlated many-body 
phases~\cite{Bloch_2008, Polkovnikov_2011,Ronzheimer,Wang,Mistakidis,Mistakidis1,Mistakidis2,Weiss,Weixu}. While tunneling in short-range interacting systems has been extensively investigated~\cite{Schneider,Bloch2005,Krutitsky2016,Weiss2009,Schafer2020}, the role of long-range interactions in quantum quenches remains largely unexplored, particularly in finite-size systems where correlations are strongest~\cite{Defenu23,Chomaz2022,Lahaye2007,Recati2023}.

Recent advances in atomic, molecular, and optical (AMO) systems have opened unprecedented opportunities to study quantum long-range interacting systems~\cite{Defenu23}. Dipolar ultracold atoms, realized experimentally with chromium~\cite{cro}, dysprosium~\cite{dys}, and erbium~\cite{erb}, exhibit long-range, anisotropic dipole–dipole interactions, leading to exotic many-body phenomena distinct from contact-interacting Bose–Einstein condensates~\cite{ref1,ref2}. These interactions stabilize unconventional quantum phases beyond the standard superfluid (SF) and Mott-insulator (MI) states, including density-wave phases, dipolar supersolids, Haldane insulators, checkerboard structures, and Mott solids~\cite{Krzysztof,Aleksandra,Boettcher:2019,Tanzi:2019,Chomaz:2019,Guo:2019,Natale:2019,Tanzi:2021}. In particular, crystallization arises from the repulsive long-range tail of the dipolar interaction, which dominates the system’s behavior and leads to strong density correlations~\cite{budha1,budha2}. Understanding the robustness of these crystal-state correlations under dynamical perturbations is a key motivation, as it provides insight into quantum correlation dynamics and informs the design of quantum simulation protocols for strongly correlated systems.

In this work, we study six dipolar bosons in a triple-well lattice with filling factor two. The prequench state is prepared in the crystal state, and the dynamics is initiated by a negative interaction quench toward SF, MI, and fermionized Mott (FMI) regimes. This protocol effectively probes the response of the system to a sudden reduction of long-range interactions to short-range ones. We employ the multiconfigurational time-dependent Hartree (MCTDH) method, as implemented in MCTDH-X, to solve the few-body Schrödinger equation. The dynamics exhibit rich intra- and inter-well tunneling, collective excitations, and fragmentation, which are characterized via density fluctuations in real and momentum space, site-resolved position variance, dynamical fragmentation, and first- and second-order Glauber correlation functions, revealing how strong initial correlations evolve under quench dynamics.

We observe that the correlations characteristic of the crystal state remain remarkably robust throughout the quench, even as the system is driven toward superfluid, Mott-insulating, and fermionized Mott regimes. This persistence occurs despite the emergence of nonequilibrium dynamics, which are predominantly governed by intrawell excitations, while interwell tunneling remains strongly suppressed.

In particular, the dynamics are dominated by local breathing and dipole-like modes within individual lattice sites, with only minimal particle transfer between wells. The negligible interwell tunneling indicates that the lattice sites remain effectively decoupled, thereby preserving the underlying long-range ordering—a hallmark of dipolar interactions—even under sudden interaction changes.

This coexistence of strong intrawell dynamics with weak interwell transport underscores the resilience of strongly correlated crystal states out of equilibrium. It further suggests that local excitations can be induced and controlled without significantly disrupting global correlation properties, providing a promising route for probing and engineering correlation-driven phenomena in finite-size ultracold lattice systems.

%We observe that the correlations of the crystal state remain remarkably robust throughout the quench, even as the system is driven toward superfluid, Mott-insulator, and fermionized Mott regimes. This persistence indicates that the long-range ordering, a hallmark of dipolar interactions, is largely immune to sudden reductions in interaction strength. Such stability highlights the inherent resilience of strongly correlated crystal states under nonequilibrium dynamics and provides a reliable framework for probing correlation-driven phenomena in finite-size lattices. Furthermore, understanding this robustness offers valuable insight for quantum simulation, as it demonstrates that key correlation features can be preserved and controlled experimentally, enabling exploration of exotic quantum phases and the design of protocols for manipulating long-range correlations in ultracold atomic systems.
%
The paper is organized as follows. In Sec. II, we present the theoretical framework, introducing the many-body Hamiltonian, the quench protocol, the methodology, and the observables used to characterize the system. In Sec. III, we discuss the results, which are subdivided into different sections to properly analyze the initial states and the quench dynamics. Finally, Sec. IV summarizes our main findings and conclusions.

\section{Theoretical framework}

\subsection{Hamiltonian and quench protocol}
The many-body Hamiltonian for $N$ dipolar bosons confined in a one-dimensional optical lattice is given by
\begin{equation}
\hat{H} = \sum_{i=1}^{N} \left[-\frac{\hbar^2}{2m} \frac{\partial^2}{\partial x_i^2} + V_0 \sin^2(k_0 x_i)\right] + \sum_{i<j} V_{\rm int}(x_i-x_j),
\end{equation}

Here, $x_i$ denotes the coordinate of the $i$-th particle. The optical lattice is characterized by its depth $V_0$ and wave vector $k_0$, and a finite number of sites with hard-wall boundary conditions is assumed. Quasi-one-dimensional confinement is ensured by tight transverse trapping. The dipolar interaction is modeled as

\begin{equation}
V_{int}(x_i-x_j) = g_s\delta(x_i-x_j) +\frac{g_d}{|x_i-x_j|^3 + \alpha_0},
\end{equation}

where $g_s$ and $g_d$ denote the short-range and dipolar coupling strengths, respectively. The parameter $\alpha_0$ encodes the transverse confinement and $g_s$ regularizes the short-distance behavior of the dipolar interaction, ensuring a finite interaction at $x_i \rightarrow x_j$ 
while preserving the asymptotic $1/|x_i-x_j|^{3}$ decay at large separations.

Prequench states are prepared in an optical lattice of depth $V_0=8 E_r$, where $E_r = \frac{\hbar^{2} k_{0}^{2}}{2m}$ is the recoil energy. The Hamiltonian is expressed in dimensionless units by rescaling energies with $\frac{\hbar^2}{mL^2}$, where $L$ is set by the lattice period.

We consider six bosons in three lattice sites ($N=6$, $S=3$), corresponding to double filling. This configuration is minimal yet sufficient to capture all relevant many-body phases—superfluid, Mott-insulating, fermionized Mott, and crystal states—while keeping the post-quench dynamics computationally tractable.

The prequench ground states serve as the initial condition for nonequilibrium dynamics, which are induced by a sudden reversal of the interaction strength. This quench protocol allows us to probe the interplay between intrawell excitations, weak interwell tunneling, and the robustness of long-range correlations under nonequilibrium conditions.

The nonequilibrium dynamics of the system are governed by the time-dependent many-body Schrödinger equation
\begin{equation}
\hat{H} \Psi(t) = i \hbar \frac{\partial \Psi(t)}{\partial t}.
\end{equation}

To investigate the dynamical response following a negative interaction quench, we consider three distinct quench protocols. The prequench state is prepared as a crystal state of six strongly interacting dipolar bosons in a triple-well lattice, with $g_d=15$
and lattice depth $V_0=8$ (Fig.~\ref{fig_initial}(d)). The post-quench interaction strengths define three target states: 1) Superfluid state: 
$g_d=0$ (Fig.~\ref{fig_initial}(a)). 2) Mott-insulator state: $g_d=0.1$ (Fig.~\ref{fig_initial}(b)). 3) 
Fragmented Mott state: $g_d=1$ (Fig.~\ref{fig_initial}(c)).

Each quench corresponds to a sudden transition from a long-range-interaction-dominated regime to one where short-range interactions govern the dynamics. The resulting post-quench states exhibit distinct intrawell excitations, weak interwell tunneling, and characteristic correlation patterns.

The quench protocol is implemented as follows. The ground state of the strongly interacting Hamiltonian ($g_d=15$)
is first obtained, representing the crystal phase. At $t=0$, the system is propagated under the Hamiltonian with the post-quench interaction strengths 
$g_d=0,0.1,1 $, while the lattice depth is kept fixed at $V_0=8$. To probe controlled interwell dynamics, we additionally consider simultaneous lattice ramping, where $V_0$ is reduced from $8$ to $4$ during the quench.

Experimentally, such interaction quenches can be realized via confinement-induced resonances~\cite{Haller2010CIR}, while the dipolar interaction strength can be tuned using Feshbach resonance techniques~\cite{interaction, Giovanazzi2002Dipolar}.

\subsection{Method}

To investigate the quench dynamics of dipolar bosons, we employ the multiconfigurational time-dependent Hartree method for indistinguishable particles (MCTDH-X)~\cite{Alon:2008,Lode:2016,Fasshauer:2016,Lin:2020,Lode:2020,MCTDHX}. The method for bosons (MCTDHB) has been extensively applied to single-species structureless bosons and is particularly well suited for capturing the nonequilibrium dynamics of interacting ultracold systems~\cite{rhombik_scirep,rhombik_shubhro_pra,Bar-Molignini:2025,rhombik_acc}. MCTDH-X has furthermore been successfully applied to a variety of dipolar-interacting systems, providing accurate access to correlation effects and time-dependent observables~\cite{Fischer:2015,budha1,chatterjee:2019,Bera:2019,chatterjee:2020,rhombik_jcp,Hughes:2023, Molignini_2025,Molignini:2022,Molignini:2024,rhombik_epjplus,rhombik_pre}.

MCTDH-X solves the time-dependent many-body Schrödinger equation using a variationally optimized, time-dependent many-body basis. The many-body wavefunction is expressed as a superposition of $M$ time-dependent permanents $|\mathbf{n};t\rangle$ with time-dependent coefficients $C_{\mathbf{n}}(t)$:
\begin{equation}
|\Psi(t)\rangle = \sum_{\mathbf{n}} C_{\mathbf{n}}(t) , |\mathbf{n};t\rangle,
\label{many_body_wf}
\end{equation}

where $|\mathbf{n};t\rangle$ denotes a many-body configuration with occupation numbers $\mathbf{n}$ distributed over the $M$ single-particle orbitals. This adaptive basis allows an efficient and accurate description of strongly correlated dynamics, including fragmentation, tunneling, and collective excitations, going beyond the limitations of static mean-field or standard Bose-Hubbard approaches~\cite{ref3,ref4,Tan2016,Streltsov2013}.

The permanents $|\mathbf{n};t\rangle$ are constructed from the $M$ time-dependent single-particle orbitals as
\begin{equation}
|\mathbf{n};t\rangle = \frac{1}{\sqrt{\prod_i n_i!}} \prod_{i=1}^{M} \left( \hat{b}_i^\dagger(t) \right)^{n_i} |{\rm vac}\rangle,
\end{equation}

where $\hat{b}_k^\dagger(t)$ creates a boson in the $i$-th orbital at time $t$, $n_i$ denotes the occupation of the $i$-th orbital, and $|vac \rangle$ is the vacuum state. The combination of time-dependent orbitals and coefficients enables the method to optimally capture evolving correlations and fragmentation during the quench.

Both the orbitals and the expansion coefficients are variationally optimized at each time step. The equations of motion are derived using the time-dependent variational principle within a Lagrangian formulation~\cite{variational1}, ensuring an optimal representation of the many-body dynamics and correlations throughout the quench.

In the limit $M=1$, MCTDH-X reduces to the Gross-Pitaevskii mean-field description, while for $M \rightarrow \infty$, the expansion becomes formally exact. In practice, the number of orbitals is finite and chosen based on convergence criteria, such as the stability of observables and negligible occupation of the highest orbital.

\subsection{Observable}

To characterize the dynamics following the interaction quench, we compute several key observables. The spatial distribution of the bosons is quantified via the one-body density
\begin{equation}
\rho(x,t) = \langle \Psi(t) | \hat{\Psi}^{\dagger}(x) \hat{\Psi}(x) | \Psi(t) \rangle,
\end{equation}
where $\Psi(t)$ is the many-body wavefunction. To probe spatially resolved dynamical fluctuations, we calculate the density fluctuation
\begin{equation}
\delta \rho(x,t) = \rho(x,t) - \langle \rho(x,t) \rangle_T,
\end{equation}
where the time-averaged density is defined as
\begin{equation}
\langle \rho(x,t) \rangle_T = \frac{1}{T} \int_0^T dt, \rho(x,t),
\end{equation}
and $T$ is the total propagation time. This quantity highlights intrawell excitations, including local breathing modes and dipole-like oscillations.

The corresponding momentum-space fluctuation is obtained via the Fourier transform of the density matrix, providing insight into coherence properties and the spread of bosons across the lattice.

To quantify interwell tunneling, we compute the site-resolved position variance, which captures the fluctuation of particle number and spatial correlations between lattice sites as

\begin{equation}
\sigma_{x,D}^2(t) = \langle \Psi(t) | \hat{x}_D^2 | \Psi(t) \rangle - \langle \Psi(t) | \hat{x}_D | \Psi(t) \rangle^2,
\end{equation}
where $\hat{x}_D$ is the position operator restricted to a given well (domain) $D$. This quantity provides insight into the intra-well dynamics, including local excitations, breathing modes, and the redistribution of density during the quench process.

Finally, one- and two-body Glauber correlation functions~\cite{Glauber:1999, rhombik_pra} are evaluated to characterize the evolving quantum correlations. 
The one-body correlation function,

\begin{equation}
g^{(1)}(x,x';t)=
\frac{
\langle \Psi(t) |
\hat{\Psi}^{\dagger}(x)\hat{\Psi}(x')
| \Psi(t) \rangle
}{
\sqrt{\rho(x,t)\rho(x',t)}
}.
\end{equation}

where $\hat{\Psi}(x)$ and $\hat{\Psi}^{\dagger}(x)$ are the bosonic field annihilation and creation operators, respectively. $g^{(1)}$
captures coherence and phase correlations between different lattice sites, while the two-body correlation function,

\begin{equation}
g^{(2)}(x,x';t)=
\frac{
\langle \Psi(t) |
\hat{\Psi}^{\dagger}(x)\hat{\Psi}^{\dagger}(x')
\hat{\Psi}(x')\hat{\Psi}(x)
| \Psi(t) \rangle
}{
\rho(x,t)\rho(x',t)
}.
\end{equation}

reveals information about particle clustering, two-particle correlations, and the emergence of density fluctuations across different length scales during the quench dynamics.

\section{Results}

\subsection{Density signature of emergent phases}

\begin{figure}[tbh]
    \centering
    \includegraphics[width=\columnwidth]{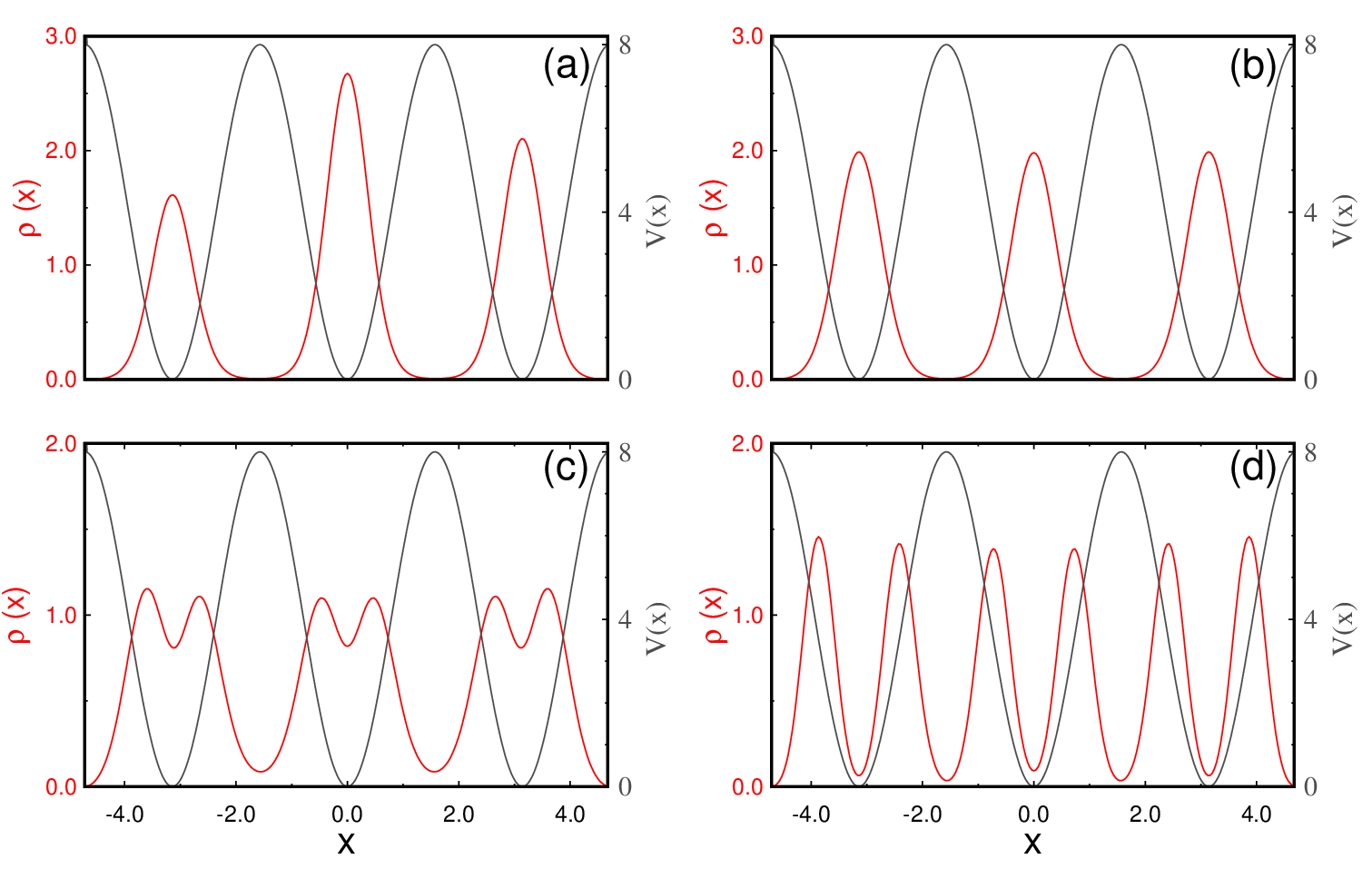}
    \caption{One-body density $\rho(x)$ (red curve) for $N=6$ bosons in $S=3$ lattice sites as a function of the dipolar interaction strength $g_d$, with lattice depth fixed at $V_0=8$. The calculations are performed using $M=24$ orbitals. (a) Superfluid (SF) phase ($g_d=0$): the density is delocalized, with a pronounced maximum in the central well due to the dominance of kinetic and lattice energies over interactions. (b) Mott-insulator (MI) phase ($g_d=0.1$): three equal density peaks across the wells indicate uniform distribution of two bosons per site and suppressed inter-well overlap. (c) Fermionized Mott (FMI) phase ($g_d=1$): the characteristic density dip within each well mimics Tonks–Girardeau–like behavior, as two bosons per site avoid spatial overlap due to strong onsite repulsion.
   (d) Crystal state (CS) ($g_d=15$): the density exhibits six well-separated maxima, corresponding to complete spatial ordering of individual bosons under dominant long-range dipolar interactions. The lattice potential is shown by the black curve. All quantities are dimensionless.
    }
    \label{fig_initial}
\end{figure}

We analyze the ground-state one-body density $\rho(x)$ as a function of the dipolar interaction strength $g_d$, keeping the lattice depth fixed at 
$V_0=8$.

For $g_d=0$ [Fig.~\ref{fig_initial}(a)], the system is fully delocalized across the triple well, corresponding to a superfluid (SF) phase. Here, the kinetic energy dominates over the interaction energy, and the imposed hard-wall boundary conditions favor enhanced population in the central well.

Increasing the interaction to $g_d=0.1$ leads to uniform redistribution of the six bosons across the three wells, with suppressed inter-well overlap, signaling the emergence of a Mott-insulator (MI) state (Fig.~\ref{fig_initial}(b)).

For stronger interactions, $g_d=1$, the particles within each well increasingly avoid spatial overlap, forming a pronounced dimer-like structure. The characteristic density dip at the center of each well indicates intrawell fermionization, and the system enters the fermionized Mott (FMI) regime [Fig.~\ref{fig_initial}(c)]. Although the short-range component of the dipolar interaction initiates this fermionization, the finite-range character of the dipole–dipole interaction enhances interparticle correlations beyond onsite repulsion, promoting stronger localization and reduced coherence.

In the strongly interacting limit $g_d=15$, the long-range $1/r^{3}$ tail dominates the many-body physics. The nonlocal repulsion energetically penalizes particle proximity not only within a single well but across the entire lattice, leading to global spatial ordering. As a result, six well-separated density maxima emerge, each corresponding to an individual boson, forming a highly correlated crystal state [Fig.~\ref{fig_initial}(d)]. Unlike the MI or FMI states, the crystal phase is a genuine manifestation of long-range interactions, characterized by strong diagonal correlations, suppressed density fluctuations, and long-range spatial ordering that cannot be captured within standard Bose–Hubbard descriptions.

This gradual evolution of $\rho(x)$ demonstrates how increasing dipolar strength drives the system from a delocalized, coherent state toward highly correlated, spatially ordered phases, providing a clear link between interaction strength and emergent many-body ordering in finite lattices.

\begin{figure}[tbh]
    \centering
    \includegraphics[width=\columnwidth]{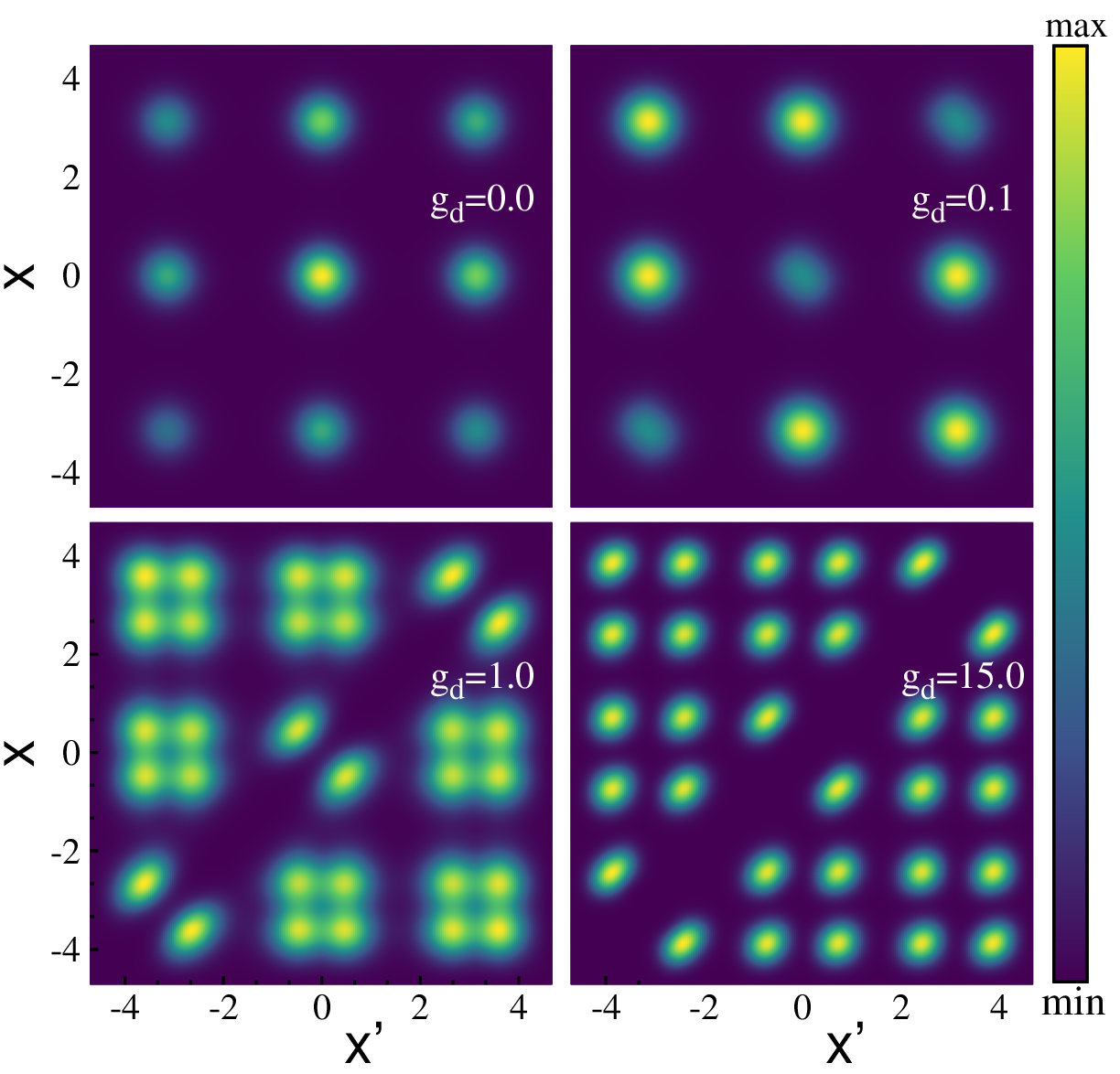}
    \caption{Two-body density $\rho^{(2)}(x,x')$ for six bosons in three lattice sites across the four emergent phases. For $g_d=0$, atoms cluster at the center with additional off-diagonal peaks due to superfluid delocalization. Increasing $g_d$ depletes the diagonal contribution $\rho^{(2)}(x,x)$, signaling enhanced correlations. At $g_d=1.0$, a correlation hole develops (fermionized Mott phase), while for $g_d=15.0$ a complete diagonal suppression indicates crystallization driven by dominant long-range interactions.
     } 
    \label{fig_initial-corr}
\end{figure}

To quantify the degree of correlations in the four emergent phases, we analyze the two-body density $\rho^{(2)}(x,x') = \langle \hat{\Psi}^{\dagger}(x) \hat{\Psi}^{\dagger}(x') \hat{\Psi}(x') \hat{\Psi}(x) \rangle $ shown in Fig.~\ref{fig_initial-corr}.

For $g_d=0.0$ [Fig.~\ref{fig_initial-corr}(a)], corresponding to the superfluid phase, the atoms predominantly cluster around the trap center $(x=x^{\prime}=0)$, reflecting strong diagonal correlations. Due to delocalization, pronounced off-diagonal peaks also appear at  ($x=0$, $x^{\prime}= \pm \pi$) and ($x= \pm \pi$, $x^{\prime}= 0$) indicating coherent inter-well correlations. 

At $g_d=0.1$ [Fig.~\ref{fig_initial-corr}(b)], in the Mott-insulator regime, particles reduce their spatial overlap. The diagonal contribution is suppressed compared to the superfluid case, and correlations become more localized within individual wells, signaling the onset of interaction-driven localization.

For $g_d=1.0$ [Fig.~\ref{fig_initial-corr}(c)], the fermionized Mott phase exhibits a pronounced correlation hole, with $\rho^{(2)}(x,x)$. The probability of finding two bosons at the same position is strongly reduced due to enhanced short-range repulsion and the nonlocal character of the dipolar interaction.

In the strongly interacting limit $g_d=15.0$ [Fig.~\ref{fig_initial-corr}(d)], long-range interactions dominate completely. The particles avoid each other over the entire lattice, leading to nearly complete suppression of diagonal correlations and the emergence of well-separated correlation peaks. The two-body density displays a crystalline pattern with a missing diagonal, reflecting strong spatial ordering and global correlation characteristic of the crystal state.

These correlation signatures provide a robust basis for exploring the dynamics of these many-body phases. By initiating the quench from a crystal state, we can probe how the inherent long-range correlations are preserved or altered when transitioning to short-range interacting post-quench states. This analysis thus forms the foundation for investigating the robustness of the crystal state’s correlations under negative interaction quenches, shedding light on the stability of exotic phases in a dynamic, nonequilibrium setting.

It is important to note that, while the 1D crystal state exhibits long-range density order, it differs fundamentally from 3D supersolids-where superfluidity and crystallization coexist due to the anisotropic dipolar interactions, supporting both off-diagonal long-range order and periodic density modulation. In contrast, the 1D crystal arises purely from strong dipolar interactions, forming a regular lattice without superfluidity, and its spatial ordering is essentially one-dimensional, lacking the multidimensional correlations present in 3D supersolids.

\subsection{Negative-Interaction-Quench Dynamics}

Building on the characterization of the ground-state properties and correlation structure of the emergent phases, we now explore the nonequilibrium dynamics induced by a negative interaction quench. Specifically, the system is initially prepared in the strongly correlated crystal state, and the dipolar interaction strength is suddenly reduced, driving the system toward regimes dominated by short-range interactions. This quench protocol enables a direct investigation of the robustness of long-range crystalline correlations and their evolution under nonequilibrium conditions.

We consider three quench protocols starting from the crystal state: (i) quenching to the superfluid (SF) phase ($g_d=0$), (ii) quenching to the Mott insulator (MI) phase ($g_d=0.1$), (iii) quenching to the fermionized Mott insulator (FMI) phase ($g_d=1.0$). These quench protocols provide a systematic framework to study the transition from a purely long-range interacting regime to states increasingly governed by short-range correlations.

%The analysis is divided into three parts: (i) tunneling dynamics, focusing on inter- and intra-well transport; (ii) correlation dynamics, utilizing one- and two-body measures to track how crystalline ordering persists or dissolves throughout the quench process; and (iii) dynamical fragmentation, examining changes in coherence and orbital occupation.
%

To comprehensively characterize the nonequilibrium dynamics following the negative interaction quenches, we monitor four key observables. First, the dynamical fragmentation quantifies the redistribution of particles among natural orbitals and provides direct insight into the evolving correlations and coherence of the system. Fragmentation is described by the time evolution of the natural occupations per particle,  $\frac{n_i}{N}$, where $n_i$ denotes the occupation of the $i$-th orbital. A nonfragmented (condensed) state has a single dominant orbital,  $\frac{n_i}{N} \simeq \mathcal O(1)$, whereas a fragmented state exhibits several significantly occupied orbitals, $\frac{n_i}{N} \simeq \mathcal O(1)$ for multiple 
$i$. In the context of interaction quenches, positive quenches from weakly to strongly correlated states enhance fragmentation, while negative quenches, such as those considered here, are expected to reduce fragmentation as correlations weaken.

The remaining three observables, previously discussed, capture complementary aspects of the dynamics: the lattice site-resolved position variance, $\sigma^{2}(t)$, quantifies interwell tunneling; real-space density fluctuation, $\delta \rho(x,t)$, reveal intrawell excitations such as local breathing and dipole-like oscillations; and the momentum-space density fluctuation, $\delta n(k,t)$, distribution highlights changes in coherence and the redistribution of particles across momentum modes. Together, these four observables provide a complete framework for analyzing the impact of negative interaction quenches on the evolution of the crystal state and its transition toward short-range interacting phases.

\subsubsection{Dynamics Following a Quench from $g_d=15$ to $g_d=0$} 

We first consider the dynamics following a sudden quench from the strongly interacting dipolar crystal ($g_d=15$) to the non-interacting limit ($g_d=0$) (superfluid regime). This represents the extreme delocalization scenario, where the crystal is expected to melt and particles are free to tunnel across the lattice. 

%To analyze the dynamics, we monitor four complementary observables: dynamical fragmentation, onsite number fluctuations (variance), real-space density fluctuations, and momentum-space density fluctuations. Together, these quantities provide a complete picture of particle redistribution, tunneling modes, and coherence properties of the system. The quench is expected to induce strong density delocalization, pronounced tunneling oscillations, and melting of the crystal, while the initially fragmented character may persist due to the strongly correlated initial state.
%

\begin{figure}[tbh]
    \centering
    \includegraphics[width=\columnwidth]{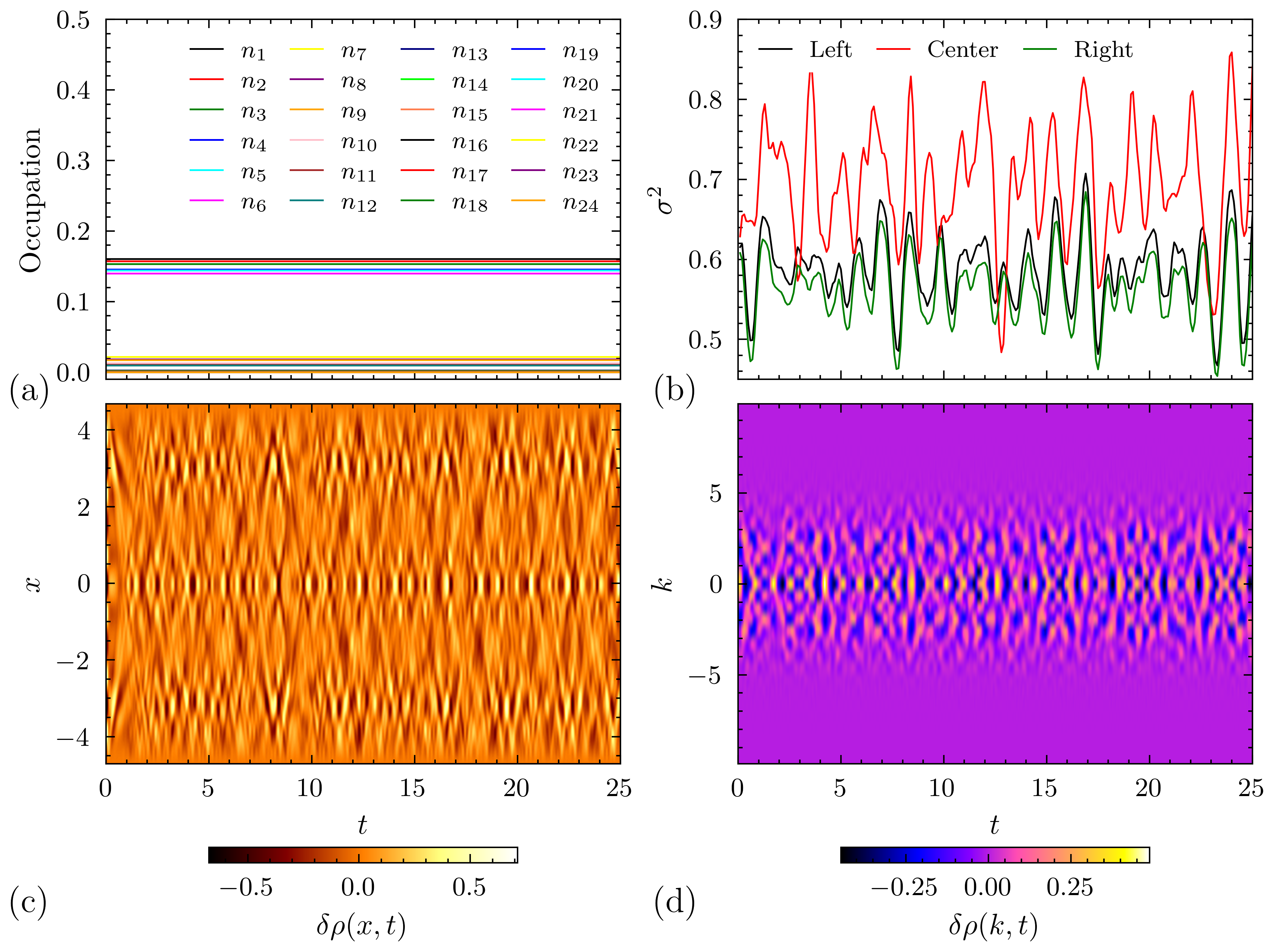}
    \caption{Nonequilibrium dynamics of six dipolar bosons in a triple-well lattice following a negative interaction quench from $g_d=15$ (crystal state) to 
$g_d=0$ (superfluid). (a) Dynamical fragmentation: time evolution of the natural occupations $n_{i}/N$ illustrating the redistribution of particles among orbitals, (b) Lattice site-resolved position variance $\sigma^{2}(t)$, (c) Real-space density fluctuations $\delta \rho(x,t)$, (d) Momentum-space fluctuations $\delta n(k,t)$. Computation is done with $M=24$ orbitals.
     } 
    \label{fig-gd=0.0}
\end{figure}

We first examine the dynamical fragmentation as presented in Fig.~\ref{fig-gd=0.0}(a). In the initial crystal state ($g_d=15$), the six lowest natural orbitals are equally populated, each with approximately $16.6\%$ of the total particle number. After the quench toward the superfluid regime ($g_d=0$), which favors a single dominant orbital, the system exhibits remarkable robustness: the initial fragmentation is largely preserved throughout the dynamics, with no significant redistribution of population among the orbitals. It indicates that the fragmentation remains frozen, highlighting that particle motion occurs within individual orbitals rather than through condensation into a coherent state.

We next present the onsite number fluctuation, $\sigma^{2}(t)$ for three lattice sites in Fig.~\ref{fig-gd=0.0}(b). It exhibits clear oscillatory behavior, reflecting tunneling-induced particle redistribution between the wells. The central well displays the largest fluctuations, ranging approximately from $0.5$ to $0.8$, while the left and right wells fluctuate with smaller amplitudes around $0.5$ to $0.6$. This indicates that the central well acts as the primary transport channel, mediating particle transfer between the edge wells. The oscillatory pattern in the central well also suggests the presence of weak intrawell excitations reminiscent of cradle or sloshing modes, arising from dipolar-mediated particle motion. Although these oscillations are moderate due to the weak tunneling and strong initial correlations, they demonstrate that even in a regime where the global fragmentation remains unchanged, local dipole-like dynamics can still emerge, highlighting the coexistence of robust long-range order with intrawell motion.

In Fig.~\ref{fig-gd=0.0}(c), we show the real-space density fluctuations, $\delta \rho(x,t)$. The fluctuations exhibit a complex spatial pattern, with significant variations both within and between wells. While the detailed structure across the three lattice sites is not easily resolved, the overall pattern indicates the presence of intrawell motion and local dipole-like excitations, consistent with the weak cradle or sloshing dynamics suggested by the variance analysis. These results highlight that, even when the global fragmentation remains robust, the system supports rich local excitations in the density profile, reflecting the interplay of long-range correlations and quench-induced dynamics.

In Fig.~\ref{fig-gd=0.0}(d), we show the momentum-space fluctuations, $\delta n(k,t)$. The fluctuations exhibit a strong central jet near $k \approx 0$, corresponding to the dominant intrawell motion in the central lattice site, consistent with the large variance observed in Fig.~\ref{fig-gd=0.0}(b). Flanking this peak are two weaker side jets, which encode contributions from the left and right wells, as seen in the real-space density fluctuations (Fig.~\ref{fig-gd=0.0}(c)). Beyond these features, $\delta n(k,t)$ is essentially flat and featureless, indicating negligible high-momentum excitations. These results show that while the global fragmentation remains robust, each lattice site contributes weak intrawell dynamics, with the central well dominating and the outer wells producing smaller, yet distinguishable, momentum-space signatures.

In summary, the negative interaction quench from $g_d=15 \rightarrow 0 $, reveals a robust yet subtly dynamic response of the 1D crystal state. Together, these four observables demonstrate that, although the quench drives the system toward a superfluid regime, the 1D crystal state retains its long-range correlations, while localized intrawell and weak interwell dynamics produce subtle, site-dependent signatures in both real and momentum space.

\subsubsection{Dynamics Following a Quench from $g_d=15$ to $g_d=0.1$} 

Building on the analysis of the negative quench to the superfluid regime, we now turn to a different scenario to further explore the robustness and dynamical response of the 1D crystal state. Specifically, we consider the sudden reduction of the interaction strength from $g_d=15$ to $g_d=0.1$, corresponding to a weakly interacting regime with characteristics of a Mott-insulator-like (MI) state. The dynamics are again analyzed through four complementary observables: the dynamical fragmentation, the onsite number fluctuations (variance), the real-space density fluctuations, and the momentum-space density fluctuations. Together, these quantities provide a comprehensive picture of particle redistribution, tunneling, and coherence in the system. The results reveal a transient phase of particle motion, followed by stabilization, highlighting the intricate interplay between tunneling, localization, and orbital occupation redistribution in the MI-like regime.

\begin{figure}[tbh]
    \centering
    \includegraphics[width=\columnwidth]{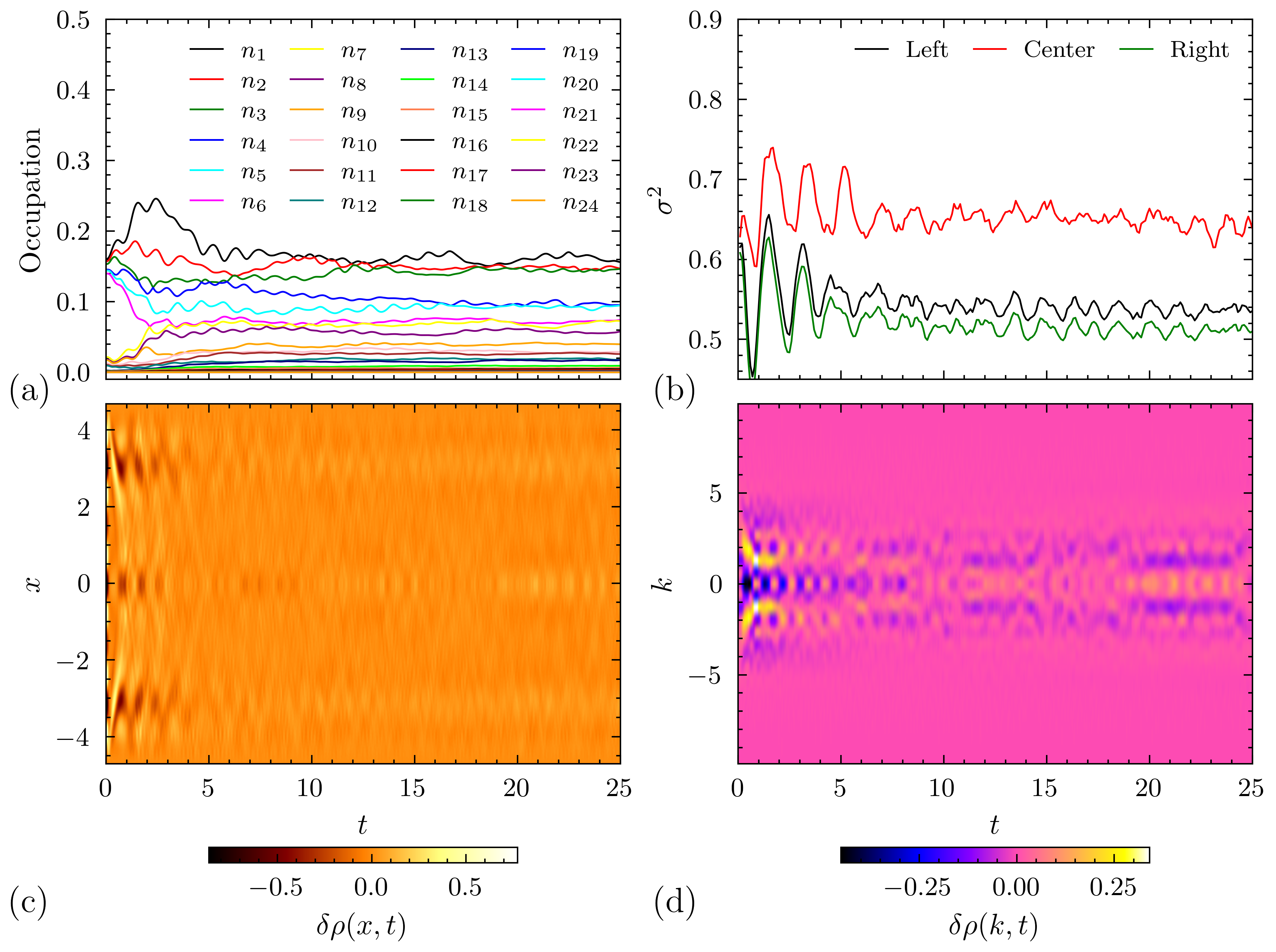}
    \caption{Nonequilibrium dynamics of six dipolar bosons in a triple-well lattice following a negative interaction quench from $g_d=15 $ (crystal state) to 
$g_d=0.1 $ (MI). (a) Dynamical fragmentation: time evolution of the natural occupations $n_{i}/N$ illustrating the redistribution of particles among orbitals, (b) Lattice site-resolved position variance $\sigma^{2}(t)$, (c) Real-space density fluctuations $\delta \rho(x,t)$,  (d) Momentum-space fluctuations $\delta n(k,t)$. Computation is done with $M=24$ orbitals.
     } 
    \label{fig-gd=0.1}
\end{figure}

In Fig.~\ref{fig-gd=0.1}(a), the dynamical fragmentation shows that the system is initially in a maximally fragmented crystal state, with the lowest six orbitals equally populated and the remaining orbitals essentially unoccupied. In contrast to the quench to $g_d=0$ (SF), where the fragmentation remained essentially frozen, here we observe a slow redistribution of population among the orbital channels, with the higher orbitals gradually gaining occupation at the expense of the initially populated six. This redistribution occurs predominantly up to $t\approx10$, after which the orbital populations stabilize and remain nearly unchanged for the remainder of the evolution. This redistribution is uniform, with no abrupt changes or anomalies, suggesting that the system maintains its fragmented character even as additional orbitals participate in the dynamics. The gradual occupation of higher orbitals reflects the system’s adjustment to the weaker interaction regime, allowing correlations to spread across more orbitals without forming a coherent condensate.

In Fig.~\ref{fig-gd=0.1}(b), we show the site-resolved variance of the particle number for each lattice well. Initially, all three wells exhibit oscillatory dynamics, reflecting transient tunneling between sites. These oscillations persist only up to $t\approx 10$, after which the fluctuations sharply decrease and settle to nearly steady values, signaling the emergence of a localized MI-like state. Specifically, the central well stabilizes near 0.65, while the left and right wells settle around 0.5 and 0.55, respectively. The slightly higher variance in the central well indicates that it continues to mediate residual particle motion, even as the system approaches quasi-stationarity. Overall, these results show that particle motion is initially active but effectively suppressed after the transient period, consistent with the gradual freezing of orbital redistribution observed in the dynamical fragmentation.

In Fig.~\ref{fig-gd=0.1}(c), we plot the real-space density fluctuations. During the early transient phase ($t \approx 10$), fluctuations are pronounced across all three lattice sites, reflecting active particle redistribution and inter-well tunneling. No single well dominates the dynamics, indicating that the transient motion is shared across the lattice. Beyond ($t >10)$, the fluctuation pattern largely disappears, leaving a featureless profile, signaling that the density redistribution halts and the system enters a quasi-stationary state. This behavior confirms that particles become effectively pinned in their respective wells, consistent with the MI-like behavior observed in the variance and fragmentation.

In Fig.~\ref{fig-gd=0.1}(d), we show the momentum-space density fluctuations $\delta n(k,t)$. During the transient phase ($t \approx 10$), distinct features appear, reflecting the short-lived redistribution of particles across the lattice. Beyond this timescale, the fluctuations in momentum space become nearly featureless, with only very weak side features corresponding to residual contributions from the three lattice sites. This behavior mirrors the real-space observations and site-resolved variance, indicating that after $t\approx 10$ the system reaches a quasi-stationary configuration. The flattening of the momentum distribution confirms that coherent particle motion is suppressed, consistent with the emergence of MI-like localization in this weakly interacting regime.

The combined analysis of Figs.~\ref{fig-gd=0.1}(a)–(d) demonstrates that the negative quench to $g_d=0.1$ induces transient particle motion, followed by a rapid stabilization of the system. Collectively, these observables reveal that the MI-like quench allows only short-lived tunneling and redistribution, after which the system enters a quasi-stationary, localized configuration with frozen dynamics, sharply contrasting with the 
dynamics observed in the SF quench.

\subsubsection{Dynamics Following a Quench from $g_d=15 $ to $g_d=1.0 $} 

To further explore the nonequilibrium response of the dipolar crystal, we consider a negative interaction quench from $g_d=15 $ to $g_d=1 $, corresponding to an intermediate regime that exhibits characteristics of a fermionized Mott-insulator-like (FMI) state. As in the previous cases, we analyze the dynamics using four complementary observables: dynamical fragmentation, site-resolved variance, real-space density fluctuations, and momentum-space density fluctuations. This quench allows us to probe how partial short-range interactions modify the particle redistribution and tunneling processes, while still preserving some of the long-range correlations of the initial crystal. The results reveal richer dynamics compared to the MI quench, reflecting stronger inter-orbital and inter-well coupling.

\begin{figure}[tbh]
    \centering
    \includegraphics[width=\columnwidth]{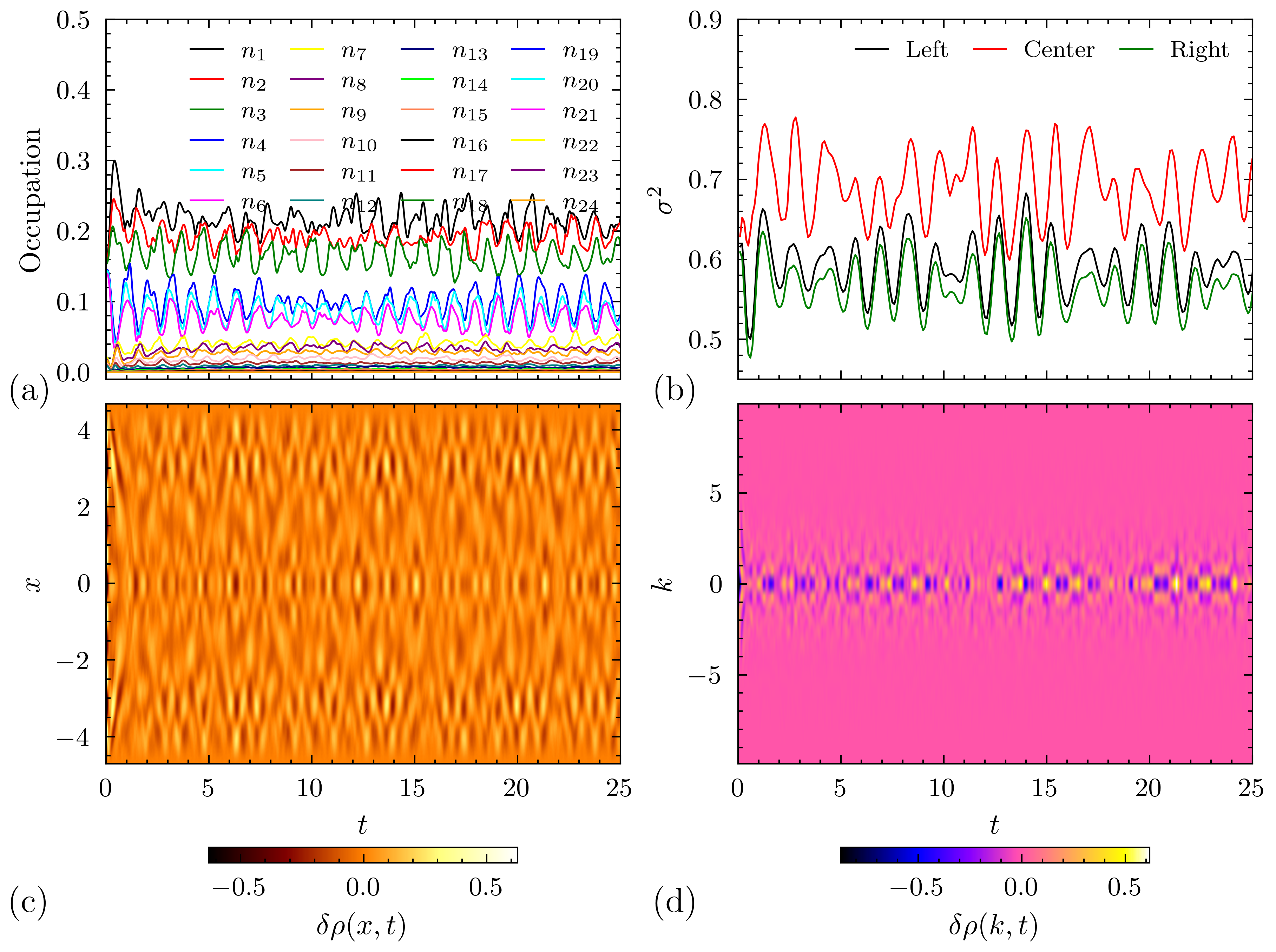}
    \caption{Nonequilibrium dynamics of six dipolar bosons in a triple-well lattice following a negative interaction quench from $g_d=15 $ (crystal state) to 
$g_d=1.0$ (fragmented MI). (a) Dynamical fragmentation: time evolution of the natural occupations $n_{i}/N$ illustrating the redistribution of particles among orbitals, (b) Lattice site-resolved position variance $\sigma^{2}(t)$, (c) Real-space density fluctuations $\delta \rho(x,t)$,  (d) Momentum-space fluctuations $\delta n(k,t)$. Computation is done with $M=24$ orbitals.
     } 
    \label{fig-gd=1.0}
\end{figure}

For the quench to $g_d=1$, the dynamical fragmentation (Fig.~\ref{fig-gd=1.0}(a)) exhibits richer behavior compared to the weaker quenches. Initially, the system is in a fully fragmented crystal state, with the lowest six orbitals equally populated ($n_i/N=1/6$) and the remaining orbitals essentially unoccupied. During the quench to $g_d=1$, these six orbitals remain the significantly occupied orbitals, but their populations fluctuate continuously throughout the dynamics, exchanging occupation with higher orbitals. The natural orbital occupations thus exhibit fluctuations across many orbitals, unlike the simpler behavior observed for weaker quenches. This indicates that the system explores a larger portion of Hilbert space, with significant multi-orbital correlations contributing to the dynamics. Despite these ongoing fluctuations, the fragmentation remains distributed among multiple orbitals, confirming that the system does not condense and retains a strongly correlated, fragmented character throughout the evolution.

This behavior contrasts with the MI quench ($g_d=0.1$), where orbital populations stabilize after $t \approx 10$, and with the SF quench ($g_d=0$), where the initial fragmentation remains essentially frozen. The FMI quench therefore highlights a richer dynamical interplay of orbital occupations, reflecting the intermediate interaction regime where long-range correlations coexist with enhanced tunneling and multi-orbital dynamics.

The site-resolved variance, [Fig~\ref{fig-gd=1.0}(b)] exhibits continuous fluctuations throughout the entire evolution, reflecting persistent inter-well tunneling and particle redistribution. The central well shows larger amplitude oscillations compared to the left and right wells, which fluctuate with similar but smaller magnitudes. This indicates that the central site acts as the primary channel for particle motion, while the edge wells participate more symmetrically. Compared to the SF quench ($g_d=0$), where fluctuations are less regular, the FMI quench exhibits more pronounced and regular oscillations, highlighting the enhanced role of multi-orbital correlations and coherent tunneling in this intermediate interaction regime. In contrast to the MI-like quench ($g_d=0.1$), where the variance stabilizes after $t \approx 10$, the sustained oscillations here indicate that the system remains dynamically active over longer timescales.

The real-space density fluctuations [Fig.~\ref{fig-gd=1.0}(c)] remain active throughout the evolution, similar to the SF quench ($g_d=0$), reflecting ongoing particle motion across the lattice. However, no clear site-resolved pattern or simple oscillatory behavior emerges, indicating that multiple tunneling modes and inter-orbital correlations interfere in a nontrivial manner. Unlike the MI-like quench ($g_d=0.1$), where fluctuations quickly settle after $t \approx 10$,  the FMI quench maintains persistent, irregular fluctuations, highlighting a dynamically active state where the system explores a larger portion of Hilbert space while retaining its fragmented character.

The momentum-space density fluctuations [Fig~\ref{fig-gd=1.0}(d)] for the FMI quench ($g_d=1)$ resemble those of the SF quench ($g_d=0$) in that fluctuations persist throughout the evolution. However, unlike the SF case, where prominent side jets appear at finite momenta, the FMI quench shows only a central feature, with the remainder of momentum space largely featureless. This indicates that particle motion is primarily centralized, with reduced coherent contributions from the edge wells. The persistent central fluctuation, reflects ongoing multi-orbital tunneling and correlations, but the absence of side jets suggests suppressed long-range coherence compared to the SF case. Overall, the momentum-space behavior reinforces the picture from variance and real-space fluctuations: the system remains dynamically active and fragmented, but with motion concentrated near the central momentum.

In summary, the FMI quench exhibits a dynamically active and strongly correlated response across all four observables. Together, these observations demonstrate that the FMI quench drives a fragmented, dynamically active state that explores a large portion of Hilbert space, in contrast to the quasi-stationary MI-like ($g_d=0.1$) quench and the more weakly fluctuating SF ($g_d=0$) quench, emphasizing the intermediate interaction regime as a bridge between frozen and fully coherent dynamics.

\subsubsection{One- and Two-Body Correlation Dynamics}

To gain deeper insight into the microscopic coherence and particle correlations beyond what is captured by variance and density fluctuations, we analyze the one- and two-body Glauber correlation functions. The one-body correlation $g^{(1)}(x,x^{\prime};t)$ provides information on the evolution of single-particle coherence across lattice sites, while the two-body correlation $g^{(2)}(x,x^{\prime};t)$ reveals how particle-particle correlations evolve during the transient dynamics. 

We focus on the quench from $g_d=15$ to $g_d=0.1$ because it drives the system from a strongly interacting crystalline state toward a weakly interacting Mott-insulator-like regime, where particle tunneling and redistribution are limited but still significant. This intermediate regime is particularly interesting because it allows us to capture transient dynamics and the formation of a quasi-stationary state; the system is no longer fully frozen. We can identify how correlations, coherence, and orbital occupations evolve during the crossover from a highly ordered crystal to a localized MI-like state, providing insight into the microscopic processes that govern relaxation, fragmentation, and the emergence of steady-state behavior in few-body dipolar systems.

\begin{figure}[tbh]
    \centering
    \includegraphics[width=\columnwidth]{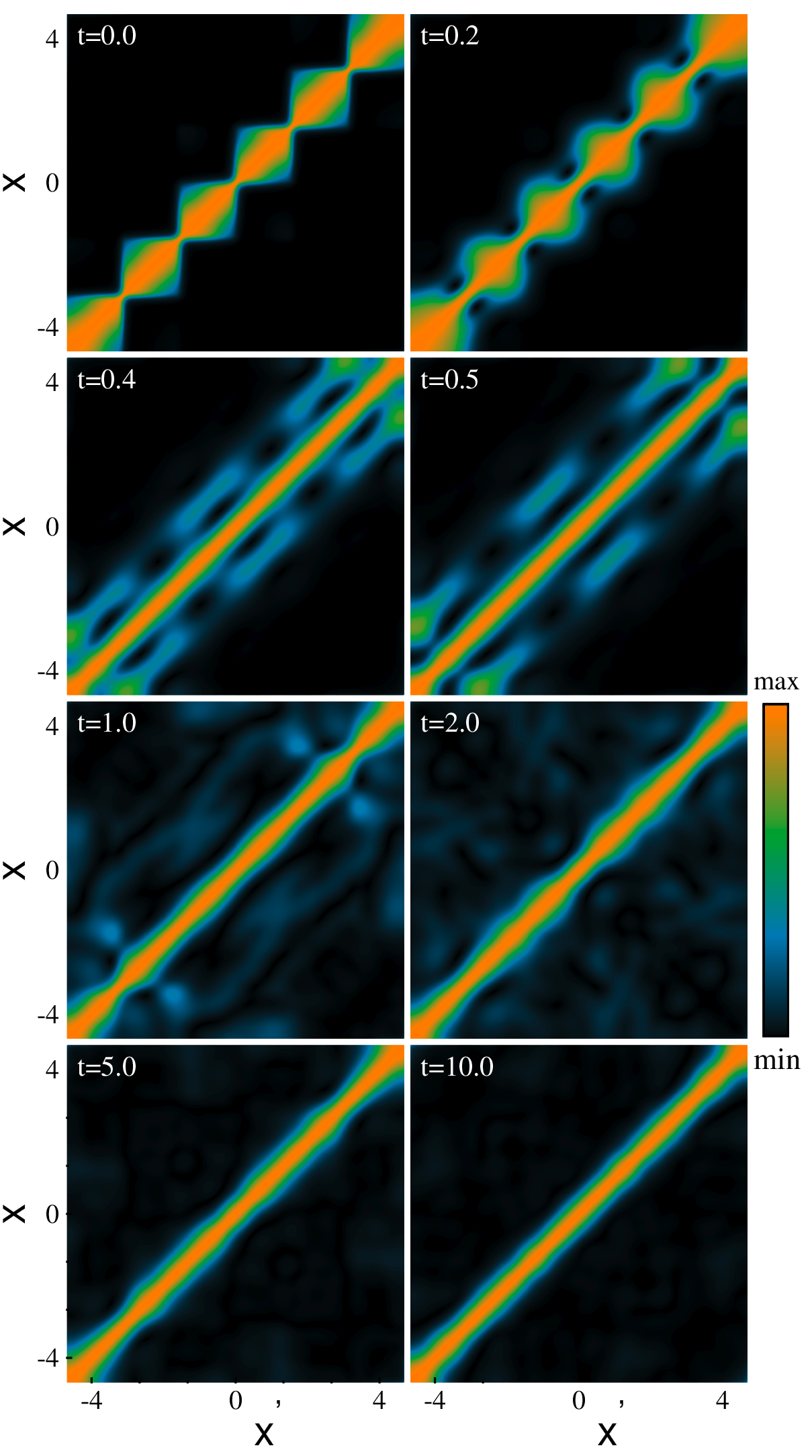}
    \caption{One-body Glauber correlation $|g^{(1)}(x,x^{\prime};t)|^{2}$ for the quench $g_d=15 \rightarrow 0.1$ at some selected time points. The growth of off-diagonal elements reflects transient tunneling and redistribution of particles, after which the coherence stabilizes, consistent with the stabilization of variance.
    }
    \label{fig-corr-1B}
\end{figure}

\begin{figure}[tbh]
    \centering
    \includegraphics[width= \columnwidth]{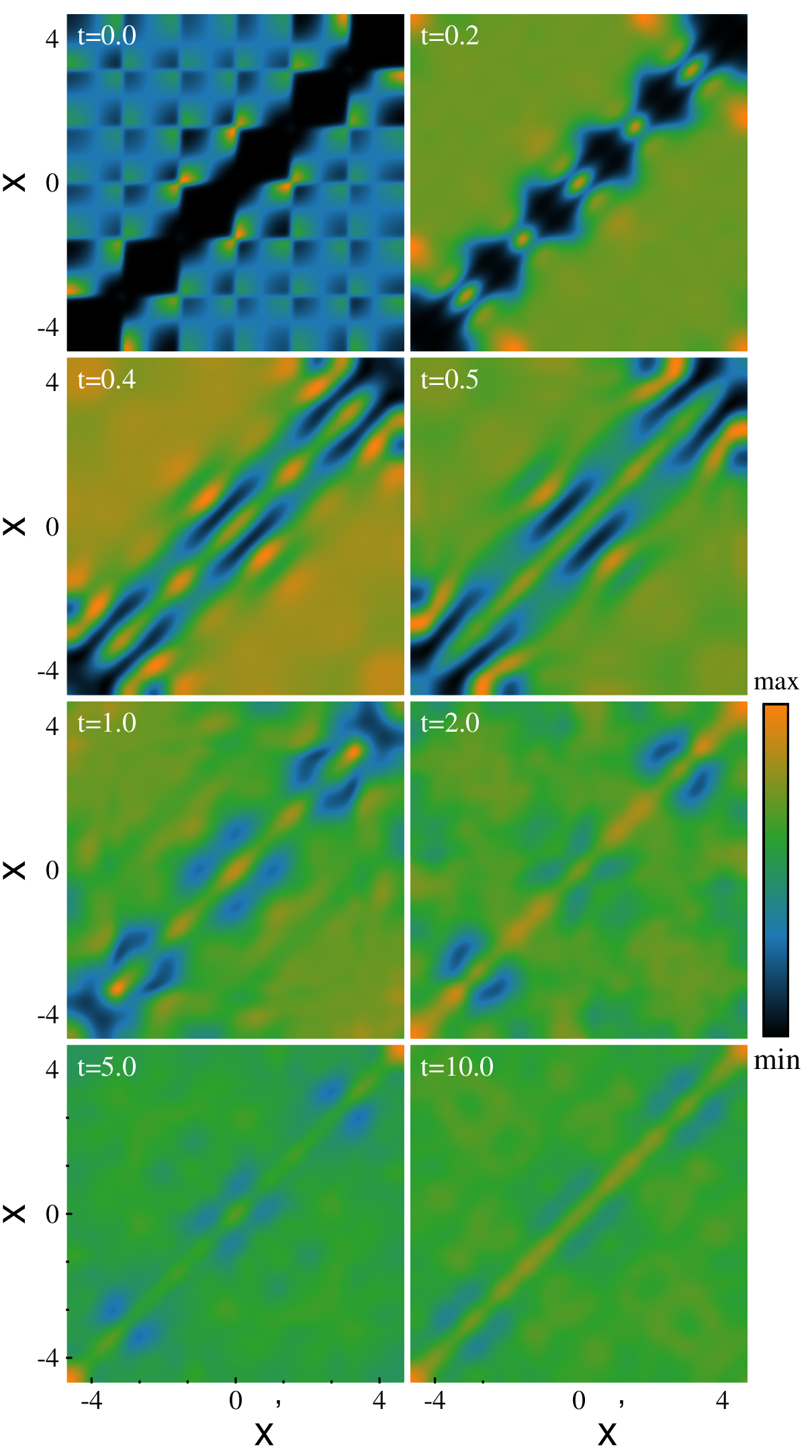}
    \caption{Two-body Glauber correlation $g^{(2)}(x,x^{\prime};t)$ for the same quench as shown in Fig.~\ref{fig-corr-1B}. Anti-correlations along the diagonal weaken transiently, indicating short-time particle redistribution, while the overall two-body correlation pattern stabilizes after $t\approx 10$, supporting the emergence of a quasi-stationary state. }
    \label{fig-corr-2B}
\end{figure}

The correlation dynamics of the MI quench ($g_d=0.1$) are presented in Fig.~\ref{fig-corr-1B} and  Fig.~\ref{fig-corr-2B}. To analyze the short-time dynamics, we examine the one- and two-body correlation functions up to the transient regime ($t \leq 10$). 
Fig.~\ref{fig-corr-1B} shows the one-body correlation function $g^{(1)}$ at selected times. In the prequench crystal state, the system is fully six-fold fragmented, and 
$g^{(1)}$ exhibits six distinct bright lobes along the diagonal ($|g^{(1)}|^{2} \simeq 1$), corresponding to six independent coherence zones. Off-diagonal correlations are completely absent (($|g^{(1)}|^{2} \simeq 0$), indicating no coherence between wells.  the crystal state becomes distorted almost immediately after the quench, with enhanced off-diagonal brightness near neighboring sites, indicating the onset of inter-well tunneling and short-range coherence. These off-diagonal contributions persist during the transient phase, reflecting particle motion and the redistribution of orbital populations, but they remain localized between adjacent wells rather than extending across the lattice.

The two-body correlation function $g^{(2)}$ is shown in Fig.~\ref{fig-corr-2B} on the same time scale. Initially ($t=0$),
the diagonal displays six correlation holes ($g^{(2)} \simeq 0$), indicating that double occupancy within a single well is highly suppressed. Immediately after the quench, the sharp structure of the correlation holes gradually fades, signaling that crystal-like localization begins to delocalize as particles start tunneling between wells. The strong antibunching effect on the diagonal diminishes over time, and by $ t \approx 10$, $g^{(2)} \approx 1$ is approximately maintained both on and off the diagonal. This evolution highlights complex inter- and intra-well tunneling dynamics, consistent with the redistribution of orbital populations observed in the dynamical fragmentation and the transient behavior seen in variance and real-space density fluctuations.

Together, the evolution of $g^{(1)}$ and $g^{(2)}$
 demonstrates that the negative interaction quench initiates a transient tunneling phase, during which particles redistribute and short-range coherence emerges, followed by a stabilization toward a quasi-stationary, weakly interacting Mott-insulator-like state. These correlation dynamics provide a microscopic confirmation of the features observed in dynamical fragmentation, site-resolved variance, and density fluctuations.

 \subsection{Quench protocol of tunneling dynamics}

Building on the previous subsections, we have seen that negative interaction quenches from a strongly correlated crystal state induce a rich variety of dynamical behavior depending on the final interaction strength. For weak post-quench interactions ($g_d=0.1$), the system rapidly localizes into a Mott-insulator-like configuration with pinned particles and minimal fluctuations. For stronger post-quench interactions ($g_d=1$),  the dynamics remain strongly correlated and fragmented, with persistent inter- and intra-well tunneling reflected in both variance and density fluctuations.

While these studies reveal the influence of interaction strength on the quench dynamics, the role of lattice confinement in shaping the redistribution and tunneling of particles remains less explored. In particular, reducing the lattice depth lowers the barrier between wells, potentially enhancing tunneling and modifying the interplay between localization and delocalization. By simultaneously quenching both the interaction strength and the lattice depth, we can probe how the competition between interparticle correlations and external confinement governs the redistribution of particles and the emergence of quasi-stationary states.

For this quench protocol, the system is initialized in the crystal state in a deep lattice with $V_0=8$ and $g_d=15$. The many-body dynamics is then investigated under a simultaneous quench, in which the interaction strength $g_d$ is abruptly reduced to smaller values while the lattice depth is suddenly ramped down to $V_0=4$.
This combined protocol allows us to probe the interplay between interaction-induced correlations and lattice confinement and to study how particle fluctuations and tunneling dynamics evolve when the system is driven toward a weakly interacting, shallow-lattice regime.

\begin{figure}[!htb]
    \centering
    \includegraphics[width= \columnwidth]{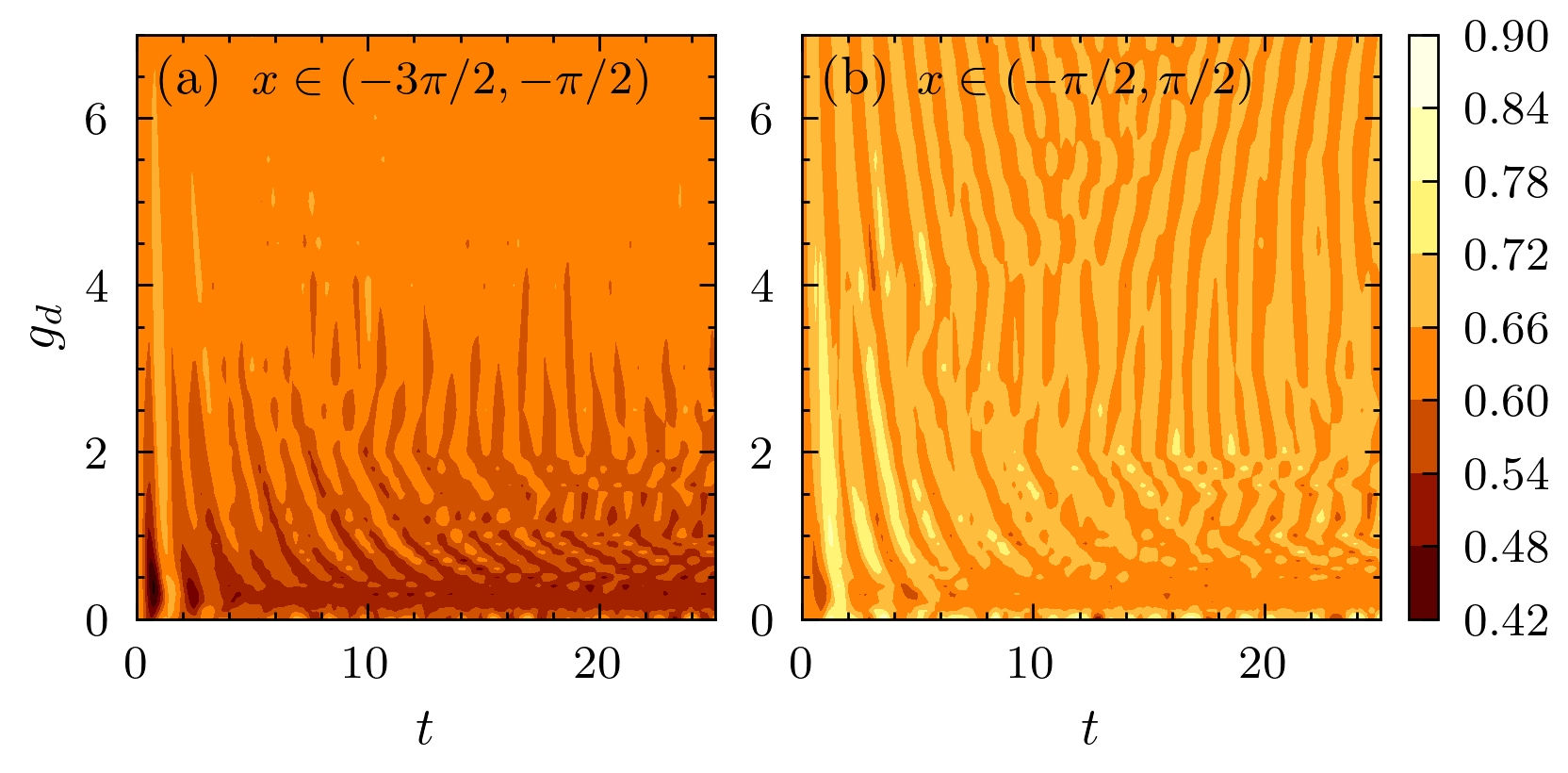}
\caption{ Dynamical regime diagram of the onsite number variance under a simultaneous quench of lattice depth and dipolar interaction strength. The system is initialized in the crystal state with $V_0=8$ and $g_d=15$, then the lattice depth is ramped to $V_0=4$ while $g_d$ is reduced to smaller values. Left panel: variance for the left (edge) well, representative of both symmetric edge sites. Right panel: variance for the central well. The color scale represents the amplitude of particle number fluctuations as a function of time and interaction strength, highlighting transient tunneling dynamics and the onset of quasi-stationary configurations. }
    \label{fig:var}
\end{figure}

To quantify the response of the system under this combined quench, we compute the onsite number variance $\sigma^{2}(t)$ for the central and left wells. Due to the left–right symmetry, the right well behaves identically to the left well. In Fig.~\ref{fig:var}, we present the dynamical regime diagram of the onsite number variance for the central and edge (left) wells as a function of $g_d$ and time. This representation allows us to visualize both the temporal evolution and the interaction-strength dependence of particle fluctuations simultaneously.

For the central well, the variance exhibits no clearly separated dynamical regions across the range of post-quench interaction strengths. This indicates that particle tunneling is continuously active, with overlapping contributions from multiple modes, and no simple or dominant tunneling pattern emerges. The persistent variance reflects the role of the central well as the main transport channel, mediating particle exchange between the edge sites throughout the dynamics.

In contrast, the left (edge) well shows two distinct regimes. For post-quench interaction strengths up to $g_d \simeq 3.5$, the variance fluctuates strongly, indicating complex particle redistribution and inter-well tunneling. This regime is dominated by diffuse motion of particles from the central site to the edges, accompanied by multiple competing tunneling frequencies, resulting in a highly nontrivial dynamical pattern. For stronger interactions ($g_d > 3.5 E_r$), the variance settles into a more regular oscillatory pattern characterized by a single dominant tunneling frequency, signaling that the dynamics have simplified and particles oscillate coherently between neighboring wells.

Overall, the diagram highlights the rich transient dynamics induced by the simultaneous quench of lattice depth and interaction strength. It demonstrates that while the central well remains a continuously active transport channel, the edge wells undergo a transition from complex multi-frequency tunneling at weak post-quench interactions to regular single-frequency oscillations at stronger interactions, illustrating the intricate dependence of particle redistribution on both lattice and interaction parameters.

\section{Conclusion}

In this work, we have investigated the nonequilibrium dynamics of a strongly interacting one-dimensional dipolar crystal under various negative interaction quench protocols. Starting from a fully fragmented crystal state in a deep triple-well lattice, we explored the system’s response to quenches toward superfluid ($g_d=0$), Mott-insulator-like ($g_d=0.1$), 
and fermionized Mott ($g_d=1$) regimes, as well as the combined quench of both interaction strength and lattice depth. The dynamics were characterized using several complementary observables: the time-dependent fragmentation, site-resolved number variance, real-space density fluctuations, and momentum-space density fluctuations, supported by one- and two-body Glauber correlation functions.

%The combined analysis of all the observables provides a unified picture of the dynamics. These observables reveal a clear separation between an early-time transient regime, %where particle redistribution and short-range coherence develop, and a later-time quasi-stationary regime, where dynamics are effectively frozen or simplified depending on the %interaction strength. Importantly, the correlation functions directly demonstrate how tunneling processes are encoded in the buildup and suppression of coherence, offering a %microscopic link between particle motion and many-body correlations.
%

A central finding of this work is the robustness of the initial crystal-state correlations, which persist even under strong reductions of the interaction strength. While the global fragmentation remains largely preserved, the system exhibits rich transient dynamics characterized by inter- and intra-well tunneling, whose nature depends sensitively on the final interaction regime. In particular, the central well consistently acts as the dominant transport channel, whereas the edge wells display a crossover from complex, multi-frequency tunneling to more regular oscillatory behavior.

Furthermore, by introducing a simultaneous quench of interaction strength and lattice depth, we have constructed a dynamical regime diagram of particle-number fluctuations. This reveals that, while the central well exhibits smooth crossover behavior due to its continuous participation in transport, the edge wells display distinct dynamical regimes, highlighting the interplay between confinement and interactions in controlling particle redistribution. These results provide insight into the nonequilibrium stability of long-range ordered states and demonstrate that key correlation features can be preserved and manipulated under quenched dynamics.

A striking observation across all quenches is the persistence of crystal-like correlations despite the sudden change in interaction strength. This robustness is significant for quantum simulation with dipolar atoms because it shows that long-range interactions can stabilize structural correlations against nonequilibrium dynamics. It suggests that few-body dipolar systems in 1D can maintain coherence of spatial patterns even when subjected to strong perturbations, making them promising platforms for investigating nonequilibrium dynamics and controlled tunneling phenomena. Furthermore, the preservation of these correlations provides a natural explanation for why fragmentation remains persistent: the system redistributes density and momentum, but the many-body orbital occupations respect the original correlation structure.

This persistence of order, together with the controllable emergence of collective modes and correlation dynamics, highlights the suitability of dipolar bosonic lattices as a versatile platform for quantum simulation of strongly correlated nonequilibrium phenomena. The ability to dynamically manipulate the interaction landscape while maintaining the integrity of the crystal order is especially promising for the controlled emulation of transport, relaxation, and driven many-body processes in long-range interacting quantum systems.

Open questions and future directions include the extension to larger lattices and higher particle numbers to explore scaling effects, the role of finite temperature and dissipation on the stability of crystal correlations, and the investigation of engineered quench protocols to selectively excite specific tunneling modes or collective oscillations such as cradle and sloshing motions. Furthermore, exploring the interplay between long-range dipolar interactions and disorder or lattice modulation could reveal new pathways for realizing dynamically tunable quantum phases.

%\%bibliographystyle{apsrev4-2}
%\bibliography{ref}

\section*{Acknowledgments} 
BC gratefully acknowledges S. I. Mistakidis for proposing the original idea of this work and for invaluable guidance in its development.

\appendix

\section{Orbital convergence}

Since the MCTDHB approach is based on a variationally optimized truncated orbital expansion, establishing numerical convergence with respect to the number of single-particle functions is crucial for ensuring the reliability of the obtained nonequilibrium many-body dynamics. In the main text, the convergence of the three considered negative interaction quenches was verified through the evolution of the dynamical fragmentation using $M=24$ orbitals. As an additional and more stringent consistency check, we analyze here the site-resolved position variance, which is highly sensitive to both local intrawell excitations and interwell tunneling processes.

\begin{figure}[htbp]
    \centering
    \includegraphics[width=\columnwidth]{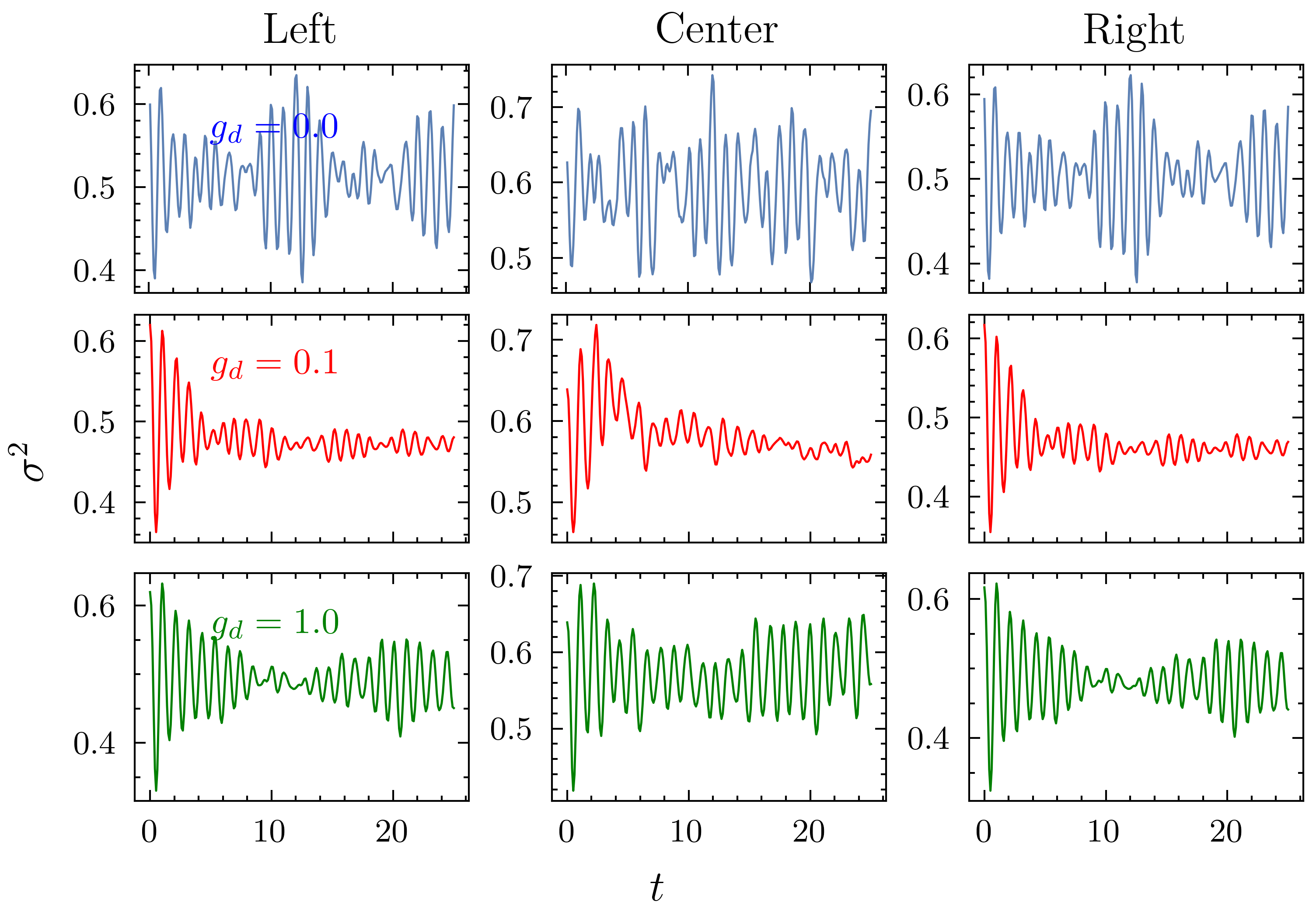}
    \caption{Time evolution of the site-resolved position variance following negative interaction quenches from the initial crystal-state configuration at $g_d=15$, obtained using $M=24$ orbitals within the MCTDHB framework. The first, second, and third columns correspond to the left, central, and right wells, respectively. The top, middle, and bottom rows show quenches to $g_d=0.0$, $g_d=0.1$, and $g_d=1.0$. The dynamics reveal the spatially resolved fluctuation response associated with local intrawell excitations and interwell tunneling processes.}
    \label{fig-M=24}
\end{figure}

\begin{figure}[htbp]
	\vskip 0.2cm
    \centering
    \includegraphics[width=\columnwidth]{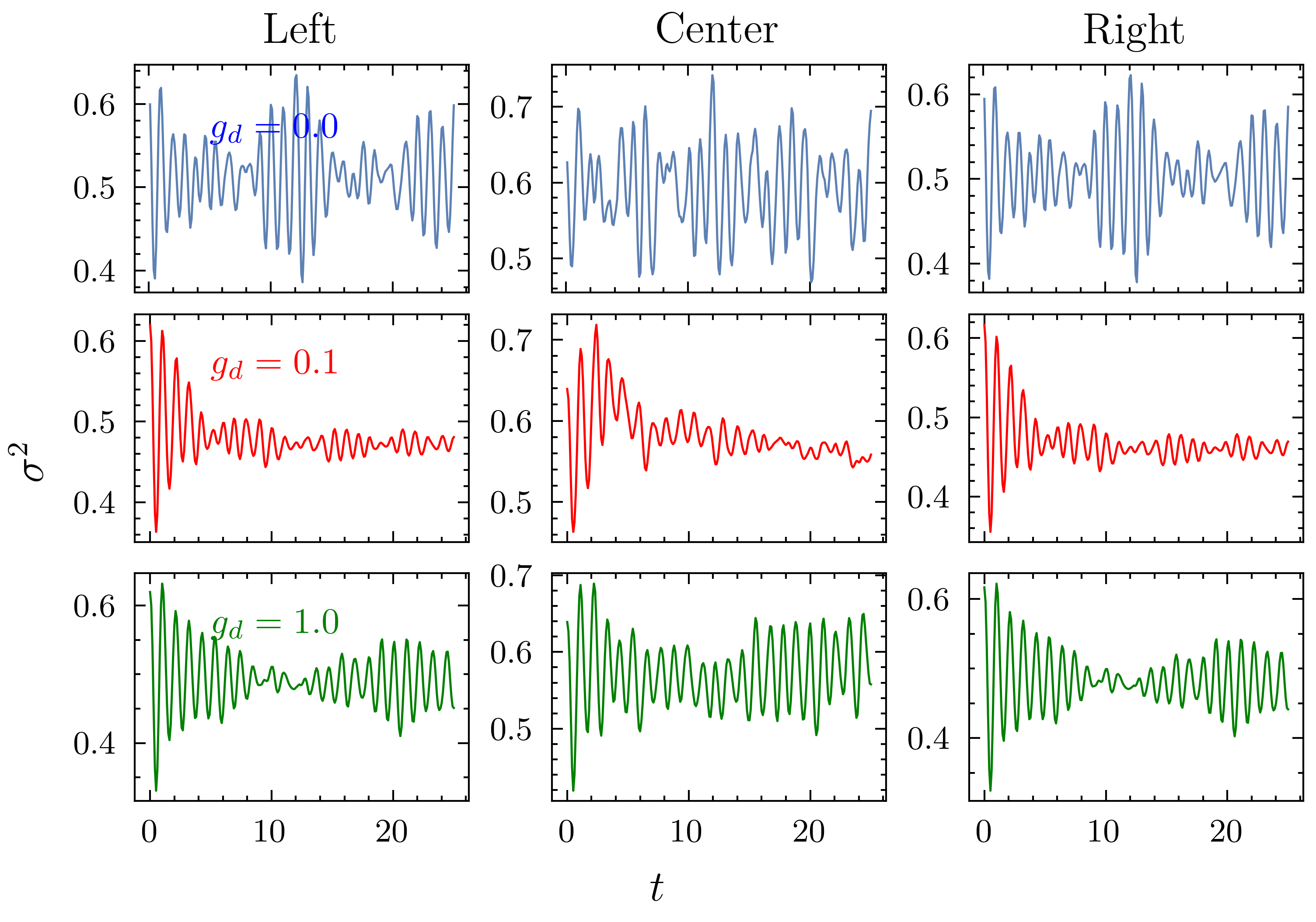}
    \caption{Same as Fig.~\ref{fig-M=24}, but for $M=27$ orbitals. The close agreement between the $M=24$ and $M=27$ results demonstrates the numerical convergence and robustness of the variance dynamics with respect to the number of orbitals employed in the MCTDHB expansion.}
    \label{fig-M=27}
\end{figure}

Figures~\ref{fig-M=24} and \ref{fig-M=27} present the corresponding variance dynamics obtained with $M=24$ and $M=27$ orbitals, respectively, following quenches from the initial crystal-state configuration at $g_d=15$. In each figure, the first, second, and third columns correspond to the left, central, and right wells, respectively, while the top, middle, and bottom rows show quenches to $g_d=0.0$, $g_d=0.1$, and $g_d=1.0$. 

For all considered quenches, the variance evolution obtained with $M=24$ and $M=27$ is found to be practically indistinguishable over the entire propagation time.
In particular, the oscillation frequencies, amplitudes, damping behavior, and temporal modulation patterns exhibit excellent agreement between the two calculations for both the edge and central wells.
The same tunneling-induced fluctuation patterns, including the enhancement or suppression of variance in specific wells depending on the final interaction strength, are reproduced identically in both calculations. Moreover, the relative phase relations between the wells and the characteristic signatures of local breathing and interwell transport remain unchanged upon increasing the orbital number from $M=24$ to $M=27$.

The absence of any visible deviations between the two calculations demonstrates that the many-body dynamics are fully converged already at $M=24$. Consequently, the variance dynamics reported in the main text reliably capture the physical response of the quenched dipolar crystal state and are not affected by truncation effects associated with the orbital basis size.

\section{Units and connection to optical-lattice experiments}\label{units}

In the MCTDHX calculations, the many-body Schr\"odinger equation is solved in dimensionless units by introducing a characteristic length scale $L$. All lengths are measured in units of $L$, while energies are expressed in units of $E_0=\frac{\hbar^2}{mL^2}$, and the corresponding unit of time is $t_0=\frac{\hbar}{E_0}=\frac{mL^2}{\hbar}$.

For an optical lattice, a natural choice of length scale is set by the laser wave vector~\cite{rr3,rr2}. For a lattice potential of the form $V_{\rm OL}(x)=V_0\sin^2(k_L x)$, with $k_L=2\pi/\lambda$, one may choose $L=1/k_L$. With this convention, the lattice period is $\pi$ in dimensionless units, while the physical lattice spacing is $d=\frac{\lambda}{2}$.

The experimentally relevant energy scale is the recoil energy, $E_R=\frac{\hbar^2 k_L^2}{2m}$. Therefore, for the choice $L=1/k_L$, the MCTDHX energy unit becomes $E_0=\frac{\hbar^2 k_L^2}{m}=2E_R$, and the corresponding time unit is $t_0=\frac{\hbar}{2E_R}$. 
This provides a direct conversion between the dimensionless parameters used in the simulations and the energy scales commonly used in optical-lattice experiments.

For example, considering $^{87}$Rb atoms in an optical lattice with wavelength $\lambda=1064\,{\rm nm}$, one obtains $E_R/h \simeq 2.03\,{\rm kHz}$, so that $E_0/h \simeq 4.06\,{\rm kHz}$, and $t_0=\frac{\hbar}{E_0}\simeq 3.9\times 10^{-5}\,{\rm s}$. 
Thus, one unit of dimensionless energy corresponds to $2E_R$, while one unit of dimensionless time corresponds to about $39\,\mu{\rm s}$ for this choice of lattice wavelength. Other experimental lattice wavelengths can be incorporated through the same recoil-energy scaling.

\bibliography{ref}

\end{document}